# Symmetries, weak symmetries, and related solutions of the Grad–Shafranov equation

G. Cicogna,[1] F. Pegoraro,[2] and F. Ceccherini[2]
[1]*Dipartimento di Fisica, "E.Fermi" dell'Università di Pisa, INFN, Sez. di Pisa, Largo B.Pontecorvo 3, I-56127 Pisa,Italy*
[2]*Dipartimento di Fisica, "E.Fermi" dell'Università di Pisa, Largo B. Pontecorvo 3, I-56127 Pisa, Italy*

We discuss a new family of solutions of the Grad–Shafranov (GS) equation that describes D-shaped toroidal plasma equilibria with sharp gradients at the plasma edge. These solutions have been derived by exploiting the continuous Lie symmetry properties of the GS equation and in particular a special type of "weak" symmetries. In addition, we review the continuous Lie symmetry properties of the GS equation and present a short but exhaustive survey of the possible choices for the arbitrary flux functions that yield GS equations admitting some continuous Lie symmetry. Particular solutions related to these symmetries are also discussed. © *2010 American Institute of Physics*.
[doi:10.1063/1.3491426]

## I. INTRODUCTION

The analysis of symmetry properties of (ordinary or partial) differential equations is a very useful tool for studying their general structure and—more importantly—for finding explicit solutions. Very schematically, we can say that the presence of a symmetry of a differential equation has two immediate applications: (i) one may obtain other solutions starting from a known one, (ii) one may look for solutions which are left invariant by this symmetry (this may be obtained considering a *reduced* equation). We will be concerned with symmetries described by continuous Lie groups of transformations: we refer, for instance, to the books[1–3] for all details about this method and for some of its typical applications.

This paper is devoted to the analysis of the Grad–Shafranov (GS) (also called Grad–Schlüter–Shafranov or Bragg–Hawthorne) equation

$$\frac{\partial^2 \psi}{\partial r^2} - \frac{1}{r}\frac{\partial \psi}{\partial r} + \frac{\partial^2 \psi}{\partial z^2} = r^2 F(\psi) + G(\psi) \qquad (1)$$

which plays a fundamental role in describing two dimensional static equilibria of magnetized plasmas under the assumption of azimuthal symmetry. Here the standard cylindrical coordinates $r, z, \varphi$ are used. In Eq. (1), $\psi(r,z)$ is the magnetic flux function such that the magnetic field can be written as

$$r\mathbf{B} = \nabla \psi \times \mathbf{e}_\varphi + I(\psi)\mathbf{e}_\varphi, \qquad (2)$$

where $\mathbf{e}_\varphi$ is the unit vector along the azimuthal direction and $I(\psi)$ is an arbitrary function of $\psi$ that is related to $G(\psi)$ in Eq. (1) by

$$G(\psi) = -I(\psi)[dI(\psi)/d\psi] \qquad (3)$$

whereas $F(\psi)$ is related to the gradient of the equilibrium plasma pressure $p(\psi)$ by

$$F(\psi) = -4\pi[dp(\psi)/d\psi]. \qquad (4)$$

When solving Eq. (1) for $\psi(r,z)$ the functions $F(\psi)$ and $G(\psi)$ must be assigned together with the boundary conditions. These may consist in assigning a boundary magnetic surface $\psi_B(r,z)$ enclosing the domain of interest in the $r-z$ plane. If this domain encloses the geometrical $r=0$ axis, additional regularity conditions at $r=0$ are required by the expression of the magnetic field in Eq. (2).

A full analysis of the symmetry properties of a larger class of partial differential equations, including GS equation as a particular case, and for generic functions $F=F(\psi)$ and $G=G(\psi)$ as well, has been performed (2006) in Ref. 4 (see also Ref. 5). A detailed analysis of the specific case of GS equation (under the name of Bragg–Hawthorne equation), with generic $F(\psi)$, $G(\psi)$, has been performed (2007) in Ref. 6.

In a more recent (2009) paper[7] some symmetry properties of the GS equation have been analyzed. This analysis actually is restricted to the case in which $F$ and $G$ are *constants*, i.e., to equilibria that are related to the Solovèv solution.[8]

In Sec. II of the present paper, referring for detailed calculations to Ref. 5 (cf. also Ref. 6), we present a short but exhaustive review of the possible choices for the functions $F(\psi)$ and $G(\psi)$ yielding GS equations which admit some continuous Lie symmetry; the main purpose is to look for possible particular solutions related to these symmetries (e.g., invariant solutions). In the two remaining sections, in addition to these "standard" Lie symmetries, we introduce the notion of "weaker" symmetries (conditional symmetries and similar) for the GS equation; the presence of such symmetries allows us to provide some other solutions to the GS equation.

In particular we find a family of solutions of the GS equation that describes D-shaped toroidal plasma equilibria and that is novel to the best of our knowledge. Similar types of D-shaped equilibria have been obtained in the literature by means of an expansion procedure in the so-called high $\beta$, large aspect ratio limit.[9,10]

## II. STANDARD LIE SYMMETRIES

For any choice of $F(\psi)$ and $G(\psi)$, Eq. (1) admits the obvious symmetry generated by $\partial/\partial z$, i.e., the translations along the $z$ axis, and the two scalings (for any real $\lambda$)

$$\psi \to \lambda\psi, \quad F \to \lambda F, \quad G \to \lambda G$$

and

$$r \to \lambda r, \quad z \to \lambda z, \quad F \to \lambda^{-4} F, \quad G \to \lambda^{-2} G.$$

### A. The nonlinear case

We consider the case in which $F$ and $G$ are nonlinear functions of $\psi$. Apart from the above trivial symmetries, the GS equation admits symmetries only with the following choices for the functions $F$, $G$:

(a) $\quad F(\psi) = \psi^{1+(2/q)}, \quad G(\psi) = \psi^{1+(1/q)}$

where $q \neq 0$ is a constant: in this case the admitted symmetry is generated by

$$X_1 = r\frac{\partial}{\partial r} + z\frac{\partial}{\partial z} - 2q\psi\frac{\partial}{\partial \psi} \quad (5)$$

(with a little but commonly accepted abuse of language, we will denote by the same symbol $X$ both the symmetry and its generator). This case may be slightly generalized: if $F(\psi)$ and $G(\psi)$ have the form

(a') $\quad F(\psi) = a(\psi+c)^{1+(2/q)}, \quad G(\psi) = b(\psi+c)^{1+(1/q)},$

where $a$, $b$, and $c$ are constants, the admitted symmetry is

$$X_1' = r\frac{\partial}{\partial r} + z\frac{\partial}{\partial z} - 2q(\psi+c)\frac{\partial}{\partial \psi}. \quad (6)$$

The particular case $q = -1/4$, i.e.,

(a'') $\quad F(\psi) = a(\psi+c)^{-7} \quad G(\psi) = b(\psi+c)^{-3}$

corresponds to an "exceptional" case for the GS equation: for this case there is the additional symmetry (see Ref. 6)

$$X'' = 2rz\frac{\partial}{\partial r} + (z^2 - r^2)\frac{\partial}{\partial z} + z(\psi+c)\frac{\partial}{\partial \psi}. \quad (7)$$

A different possibility is the following:

(b) $\quad F(\psi) = a \exp(2c\psi), \quad G(\psi) = b \exp(c\psi)$

with $c \neq 0$ and with symmetry

$$X_2 = r\frac{\partial}{\partial r} + z\frac{\partial}{\partial z} - \frac{2}{c}\frac{\partial}{\partial \psi}. \quad (8)$$

The cases $F(\psi) = 0$ or $G(\psi) = 0$ do not admit any additional symmetry; this implies that the above classification includes also the cases $a = 0$ or $b = 0$.

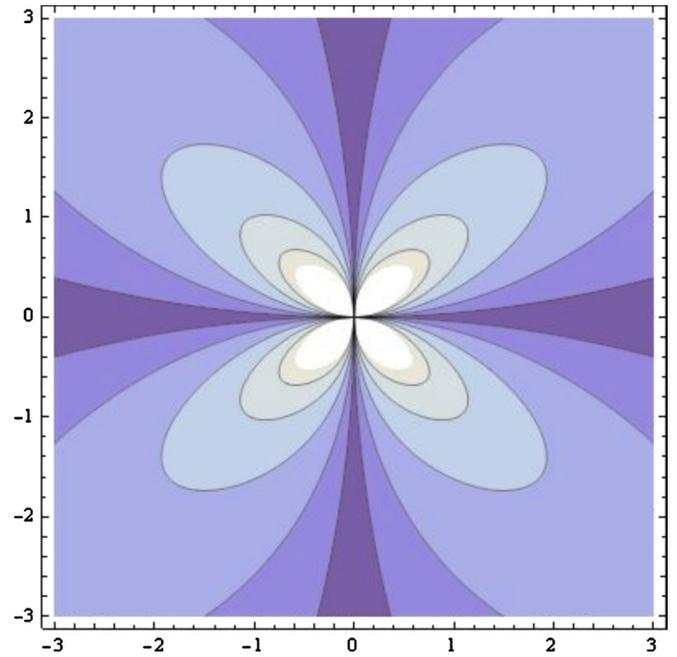

FIG. 1. (Color online) Contour plot of flux function $\psi(r,z)$ [see Eqs. (9) and (10)]. The coordinate $r$ is along the horizontal axis. For convenience the contour is shown even for $r < 0$, this clearly corresponds to the choice $\varphi = \pi$.

We now discuss some results which may be deduced from the existence of the above symmetries. The presence of the symmetry (6) in case (a') implies that if $\psi(r,z)$ is a solution to Eq. (1), then also

$$\tilde{\psi}(r,z) = \exp(2\lambda q)\psi(\exp(\lambda)r, \exp(\lambda)z) + c[\exp(2\lambda q) - 1]$$

solves the equation for any real $\lambda$.

One can look for those solutions which are left *invariant* under this symmetry: assuming for simplicity $c = 0$, these solutions have the form

$$\psi(r,z) = r^{-2q}w(r/z). \quad (9)$$

Setting $y = r/z$, the function $w(y)$ satisfies the *ordinary* differential equation (where clearly $w' = dw/dy$, etc.)

$$w''(y^2 + y^4) + w'(2y^3 - 4qy - y) + 4q(q+1)w$$
$$= aw^{1+(2/q)} + bw^{1+(1/q)} \quad (10)$$

which—as expected—involves only the symmetry-invariant variable $y$, but it is quite difficult to find its general solutions. If we look for the behavior of these solutions for $y \ll 1$, i.e., $r \ll |z|$, assuming $q > 0$ and neglecting higher order terms, we find $w \sim y^{2q}$ and $w \sim y^{2q+2}$, or

$$\psi \sim \frac{1}{z^{2q}} \quad \text{and} \quad \psi \sim \frac{r^2}{z^{2q+2}}.$$

Choosing the second possibility, the result of a numerical integration of Eq. (10) with $q = 1$, $a = -1$, $b = 1$ is given in Fig. 1 which shows a magnetic configuration with conical lobes that is singular at $r = z = 0$. This configuration has been obtained by setting $w(y) = y^{2q+2}f(y)$ and choosing $f(0) = 1$, $f'(0) = 0$.

An elementary example of invariant solution under the symmetry (6), if $F=a$, $G=b\psi^{1/2}$, which belongs to case $(a')$ with $q=-2$, $c=0$, can be given by the "cylindrical" solution $\psi=A^2 r^4$, provided that $A$ solves the equation $8A^2=a+bA$. This solution corresponds (for $a>0$, $b<0$, and $A>0$) to a constant-pitch magnetic field configuration where both $B_z$ and $B_\varphi$ are proportional to $r^2$ while the plasma pressure decreases as $(r_0^4 - r^4)$, with $r_0$ the radius of the external boundary. See Ref. 4 for additional comments on this solution.

Actually, some nontrivial particular solutions of Eq. (10) will be provided later.

We now consider the symmetry $X''$ [Eq. (7)] of the "exceptional" case $(a'')$: it implies that if $\psi(r,z)$ is a solution of the GS equation, then also

$$\tilde\psi(r,z) = [C(r,z,\lambda)]^{1/2}\psi[\tilde r(r,z,\lambda),\tilde z(r,z,\lambda)], \quad (11)$$

where

$$C(r,z,\lambda) = 1 + \lambda^2(r^2+z^2) + 2\lambda z,$$

$$\tilde r = r[C(r,z,\lambda)]^{-1} \quad \text{and} \quad \tilde z = [z+\lambda(r^2+z^2)][C(r,z,\lambda)]^{-1}$$

is a solution of the equation (for $q=-1/4$ and any real $\lambda$). This result will be important in the following.

In the same case $(a'')$, the invariant solutions under $X''$ have the form

$$\psi(r,z) = \sqrt{r}\, w(y) \quad \text{where} \quad y = \frac{r}{r^2+z^2}$$

and $w(y)$ satisfies the equation

$$w'' y^2 - \frac{3}{4} w = \frac{a}{w^7} + \frac{b}{w^3}.$$

As in the case above, a simple solution is $\psi = A\sqrt{r}$ with the condition $3A^8 + 4a + 4bA^4 = 0$.

Finally, in the case $(b)$, the presence of the symmetry $X_2$ [Eq. (8)] (with $c=1$) ensures that if $\psi(r,z)$ is a solution, then the same is true for

$$\tilde\psi(r,z) = \psi[r\,\exp(\lambda), z\,\exp(\lambda)] + 2\lambda$$

for any real $\lambda$. The solutions which are invariant under this symmetry have the form

$$\psi = -2\log r + w(r/z),$$

where now $w=w(y)$, with $y=r/z$, satisfies

$$w''(y^2+y^4) + w'(-y+2y^3) + 4 = a\exp(2w) + b\exp(w).$$

A simple cylindrical solution, as above, which is invariant under the symmetry (8), is given by $\psi = -2\log r$, if $F=a\exp(2\psi)$, $G=b\exp(\psi)$, with the condition $a+b=4$. This solution yields a configuration where both the $z$ and $\varphi$ components of the magnetic field scale as $r^{-2}$ and thus must be restricted to a domain with $r>r_0$ while the pressure difference (taken with respect to its constant value at $r=\infty$) scales as $r^{-4}$.

**B. The linear case**

We now have to consider in some detail the case in which both $F$ and $G$ are linear functions of $\psi$, i.e.,

(c)   $F(\psi) = a_0 + a_1\psi$,

$$G(\psi) = b_0 + b_1\psi \quad (a_1, b_1 \text{ not both zero}).$$

We examine the various subcases.[11]

Let us start with

$(c')$   $a_0 = b_0 = 0$

then the GS equation becomes the *linear* equation

$$\frac{\partial^2 \psi}{\partial r^2} - \frac{1}{r}\frac{\partial \psi}{\partial r} + \frac{\partial^2 \psi}{\partial z^2} = r^2 a_1 \psi + b_1 \psi. \quad (12)$$

We look for solutions of the form

$$\psi(r,z) = R(r)Z(z)$$

then the equation for $Z(z)$, $Z''=hZ$, is easily solved for any value of the arbitrary constant $h$, and $R(r)$ satisfies the equation

$$R'' - \frac{1}{r}R' + \mu R = a_1 r^2 R \quad \text{where} \quad \mu = h - b_1. \quad (13)$$

If $a_1 = \alpha^2 > 0$, the general solution for $R(r)$ is

$$R(r) = \exp(-\alpha r^2/2)\left[ c_1 U\!\left(-\frac{\mu}{4\alpha}, 0, \alpha r^2\right) \right.$$
$$\left. + c_2 L\!\left(\frac{\mu}{4\alpha}, -1, \alpha r^2\right) \right],$$

where $U$ is a confluent hypergeometric function, $L$ is a generalized Laguerre function, and $c_1$, $c_2$ are two arbitrary constants. If $\mu = a_1 = 1$ this is a combination of a function exponentially going to zero with an exponentially divergent function. If $a_1 = -\alpha^2 < 0$, the solution can be written as a combination of the real and the imaginary parts $R_1(r)$ and $R_2(r)$, both bounded and oscillating functions, of the function

$$\exp(i\alpha r^2/2) U\!\left(-\frac{i\mu}{4\alpha}, 0, -i\alpha r^2\right),$$

where $U$ is a confluent hypergeometric function with imaginary arguments.

Assuming $h<0$ one clearly gets oscillating solutions also for the function $Z(z)$, see Fig. 2 which shows a "checkerboard" of toroidal magnetic configurations.

If $a_1 = 0$, then

$$R(r) = r[c_1 J_1(\sqrt{\mu}\, r) + c_2 Y_1(\sqrt{\mu}\, r)], \quad (14)$$

where $J_1$ and $Y_1$ are Bessel functions.

Notice that, choosing in particular $h=b_1$ (i.e., $\mu=0$), Eq. (13) admits simpler solutions: assuming e.g., $h=-\nu^2<0$ and $a_1 = -\alpha^2 < 0$ one finds the solutions for $\psi(r,z)$

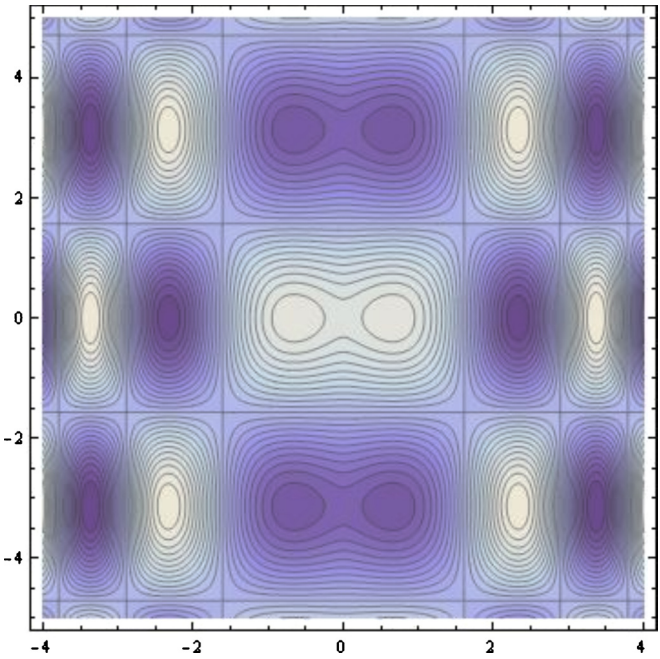

FIG. 2. (Color online) Contour plot of the flux function $\psi(r,z) = R_1(r)\cos z$ for the linear case with $\mu = \alpha = 1$. On the horizontal axis $r < 0$ corresponds to $\varphi = \pi$.

$$\psi(r,z) = \left[ c_1 \sin\left(\frac{\alpha r^2}{2}\right) + c_2 \cos\left(\frac{\alpha r^2}{2}\right) \right] \times [c_3 \sin(\nu z) + c_4 \cos(\nu z)],$$

where $c_1, \ldots, c_4$ are arbitrary constants. If $a_1 > 0$ and/or $h > 0$, the solution is similar, apart from the substitution of the functions sin, cos with sinh, cosh.

From the point of view of the symmetry properties, the GS Eq. (12) exhibits the standard symmetries of any linear equation. These simply amount to saying that any linear combination of solutions is still a solution. If $a_0$ and $b_0$ are not both zero, the situation is almost similar. Precisely, the symmetry properties of the equation imply the following:

(i) if $\psi_0(r,z)$ is any solution to the GS equation with $a_0$ and $b_0$ not both zero, then also

$$\psi_1(r,z) = \psi_0(r,z) + w_1(r,z) \qquad (15)$$

solves the same equation for any function $w_1(r,z)$ satisfying the linear GS Eq. (12);

(ii) the same holds (but this is clearly less useful in practice) for the function

$$\psi_2(r,z) = \lambda \psi_0(r,z) + w_2(r,z)$$

for any $\lambda$, if $w_2(r,z)$ satisfies this other GS equation

$$\frac{\partial^2 w_2}{\partial r^2} - \frac{1}{r}\frac{\partial w_2}{\partial r} + \frac{\partial^2 w_2}{\partial z^2} = r^2(a_0(1-\lambda) + a_1 w_2) + b_0(1-\lambda) + b_1 w_2.$$

As an application of the property (i) above, let (see Ref. 12)

$(c'')$  $a_1 = 0$,  $b_0 = 0$  $b_1 \neq 0$

then $F(\psi) = a_0$, $G(\psi) = b_1\psi$. In this case an obvious solution $\psi_0(r,z)$ to the GS equation is

$$\psi_0(r,z) = -\frac{a_0}{b_1}r^2$$

which corresponds to a uniform magnetic field component along $z$, while $w_1(r,z)$ solves Eq. (12) with $a_1 = 0$. Writing $w_1(r,z) = R(r)Z(z)$ as before, $R(r)$ is given by a combination of Bessel functions, see Eq. (14).

If instead

$(c''')$  $a_0 = 0$,  $b_1 = 0$,  $a_1 \neq 0$

there is a more complicated particular solution: if, e.g., $a_1 = -\alpha^2 < 0$ one has

$$\psi_0(r,z) = \frac{b_0}{2\alpha}\left[ \sin\left(\frac{\alpha r^2}{2}\right)\text{Ci}\left(\frac{\alpha r^2}{2}\right) - \sin\left(\frac{\alpha r^2}{2}\right)\text{Si}\left(\frac{\alpha r^2}{2}\right) \right],$$

where Si and Ci denote the functions sine integral and cosine integral, and the remaining term $w_1(r,z)$ solves the linear GS Eq. (12) as before.

In the cases

$(c^{iv})$  $a_0 = 0$,  or  $b_0 = 0$

i.e., $F = a_1\psi$, $G = b_0 + b_1\psi$ or $F = a_0 + a_1\psi$, $G = b_1\psi$, finding even one particular solution $\psi_0$ is a much more difficult task. With $a_0 = a_1 = -1$, $b_0 = 0$, $b_1 = -2$ a numerical integration of the equation shows that the solution is actually very similar to the solutions obtained in the above cases $(c')$, $(c'')$, $(c''')$.

Finally, the remaining possibility

(d)  $F(\psi) = a_0$,  $G(\psi) = b_0$

which corresponds to the Solovèv case, has been considered in full detail in Ref. 7; see also Ref. 13.

Up to a possible substitution of $\psi$ with $\psi + c$, the cases $(c'), \ldots, (c^{iv})$ and (d) cover all possible forms of the GS equation when $F$ and $G$ are linear functions, $F = a_0 + a_1\psi$, $G = b_0 + b_1\psi$.

## III. CONDITIONAL SYMMETRIES

The classification $(a), \ldots, (d)$ examined in the previous section exhausts all the possible Lie symmetries admitted by GS equations.

One can then try to look for weaker notions of symmetries, and in particular for the existence of *conditional symmetries*[14–16] (for a fairly complete list of papers devoted to this notion and some examples of its applications, see, e.g., Refs. 17–19).

Let us recall that a conditional symmetry $Y$ generates a transformation which does *not* map solutions into solutions, but defines some $Y$-invariant variables with the property that the solutions of the equation once written in terms of these variables are also solutions of the initial equation. In general, finding in a systematic way the conditional symmetries of a differential equation may be very difficult (especially when for *any* choice of $F(\psi)$. This symmetry can be useful because, introducing the symmetry-invariant variable

$$s = \tfrac{1}{2}r^2 - \kappa z$$

the GS equation is transformed into the very simple reduced *ordinary* differential equation

$$\psi_{ss} = F(\psi)$$

which can be easily integrated or reduced to quadratures. These solutions however lead to magnetic fields that are singular at $r=0$. For instance, if $F(\psi) = -1/\psi^3$, a solution to the GS equation is

$$\psi(r,z) = (r^2 - 2\kappa z + c)^{1/2}$$

or, if $F(\psi) = \exp(\psi)$, a solution is

$$\psi(r,z) = \log[8c^2 \operatorname{cosech}^2(cr^2 - 2c\kappa z + c_0)]$$

for any constants $c$, $c_0$.

(b) if $F(\psi)=0$ and for *any* $G(\psi)$, a conditional symmetry is the rotation symmetry

$$Y = z\frac{\partial}{\partial r} - r\frac{\partial}{\partial z}. \quad (17)$$

In terms of the invariant variable $s = r^2 + z^2$ the GS equation becomes

$$4s\psi_{ss} + 2\psi_s = G(\psi).$$

If, e.g.,

$$G(\psi) = b\psi^\beta,$$

where $b$ and $\beta \neq \pm 1$ are real constants, the GS equation admits then the solution

$$\psi(r,z) = A(r^2 + z^2)^\gamma \quad \text{where} \quad \gamma = (1-\beta)^{-1},$$

$$A = \left(\frac{b}{4\gamma^2 - 2\gamma}\right)^\gamma. \quad (18)$$

Again, these solutions lead to magnetic fields that are singular at $r=0$.

## IV. "WEAK" CONDITIONAL SYMMETRIES. THE D-SHAPED EQUILIBRIUM SOLUTION

It is also possible to introduce conditional symmetries of even weaker type (see Ref. 16, 17, and 19). Without entering into technical details, let us consider this example

$$Y = z\frac{\partial}{\partial r} - \sigma r\frac{\partial}{\partial z}, \quad (19)$$

where $\sigma \neq 0$ and $\sigma \neq 1$ may be either a positive or a negative constant. Writing the GS equation in terms of the $Y$-invariant variable

$$s = \sigma r^2 + z^2$$

one obtains

$$4[(\sigma^2 - \sigma)r^2 + s]\psi_{ss} + 2\psi_s = r^2 F(\psi) + G(\psi)$$

which does contain $r$ together with $s$ (in this equation the variable $r$ plays the role of a parameter, and its presence

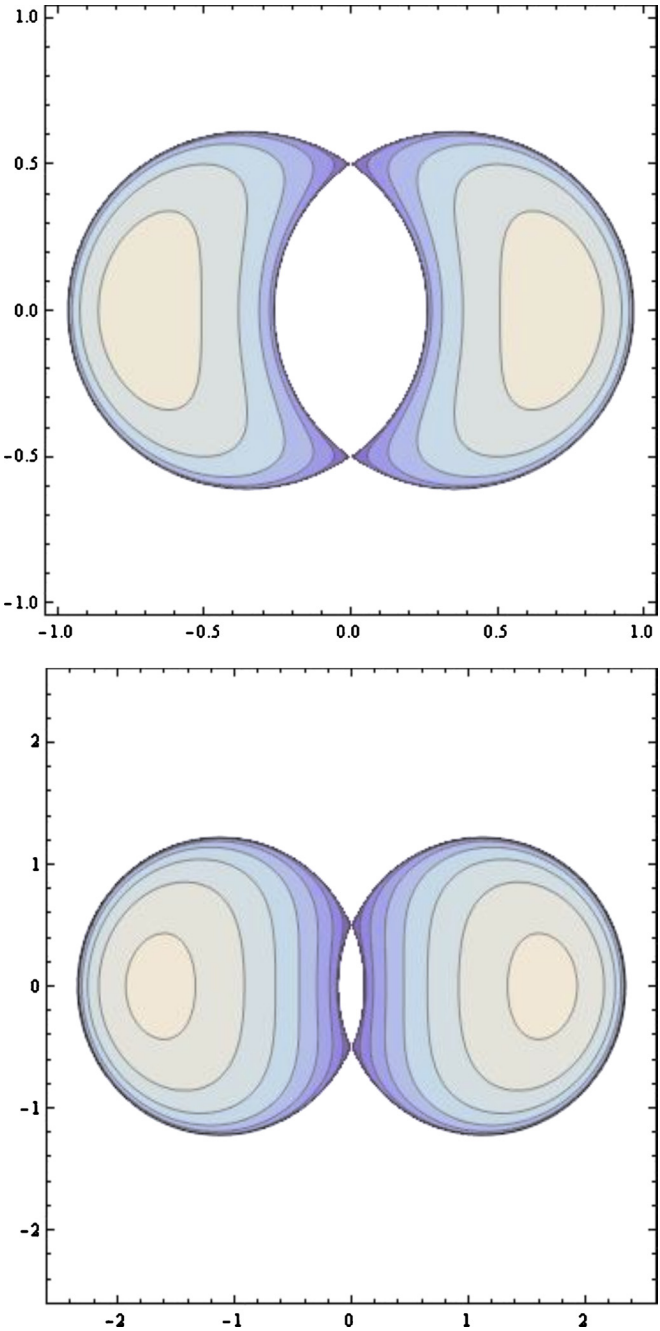

FIG. 3. (Color online) Counterplots of the flux function $\psi(r,z)$ for the D-shaped solutions in Eq. (24) corresponding to $\lambda=1$, $A=-1$, $\sigma=-0.5$, left frame, and to $\sigma=-5$, right frame. On the horizontal axis $r<0$ corresponds to $\varphi=\pi$.

the equation contains derivatives of order higher than one, cf. Ref. 14).

There are two interesting conditional symmetries for some particular cases of the GS equation.

(a) in the case

$$G(\psi) = \kappa^2 F(\psi) \quad \text{with} \quad \kappa = \text{const}$$

the admitted conditional symmetry is

$$Y = \kappa\frac{\partial}{\partial r} + r\frac{\partial}{\partial z} \quad (16)$$

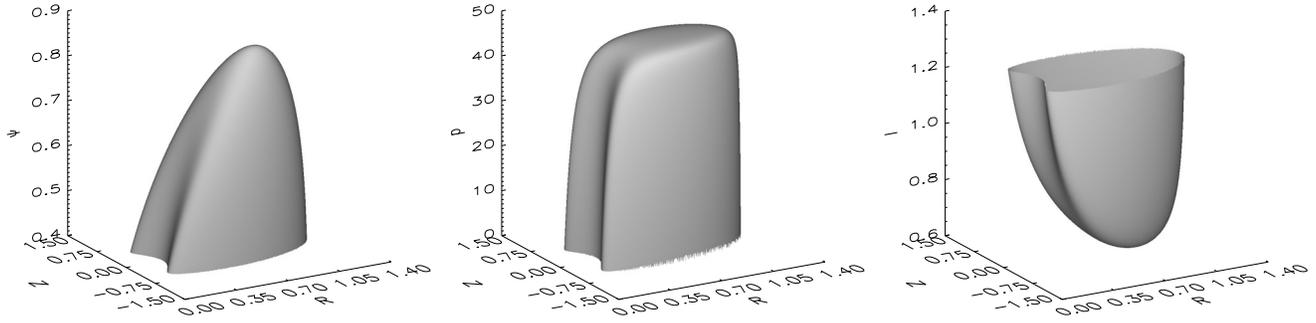

FIG. 4. 3D plots of the flux function $\psi(r,z)$ (left frame) of the pressure $p(\psi)=p(r,z)$ (center frame) and of the function $I(\psi)=I(r,z)$, that gives the azimuthal component of the magnetic field times $r$, (right frame) for the D-shaped solutions in Eq. (24) corresponding to $\lambda=1$, $\sigma=-1$. All functions have been restricted to the domain $\psi > \psi_0 = 0.4$, where $\psi = \psi_0$ represents the plasma boundary. The integration constant for the function $p(\psi)$ has been fixed such that the pressure vanishes at the plasma boundary. Contrary to previous figures only positive values of $r$ are shown.

indicates that $Y$ is not a proper conditional symmetry), but can be naturally split into *two* ordinary differential equations involving only $s$ (as happens in the case of weak conditional symmetries[17])

$$4(\sigma^2 - \sigma)\psi_{ss} = F(\psi) \quad \text{and} \quad 4s\psi_{ss} + 2\psi_s = G(\psi). \quad (20)$$

Clearly, these equations admit solutions only if there is a precise relationship between $F(\psi)$ and $G(\psi)$. Solutions of these equations can be found of the form

$$\psi(s) = (As)^{-q} = A^{-q}(\sigma r^2 + z^2)^{-q} \quad (21)$$

for any real $q \neq 0$, with

$$F(\psi) = a\psi^{1+(2/q)}, \quad G(\psi) = b\psi^{1+(1/q)}, \quad (22)$$

where the constants $A, a, b$ are related by

$$a = 4A^2(\sigma^2 - \sigma)q(q+1), \quad b = 2Aq(2q+1). \quad (23)$$

Before considering explicit examples of this case, let us emphasize that it falls precisely in case $(a')$ of our symmetry classification in Sec. II (with $c=0$): this explains why we use the notation $q$ for the exponent in Eq. (21). In addition, it is easy to verify that the solutions (22) turn out to be invariant under the symmetry $X_1$ [Eq. (5)]; accordingly, they have the form $r^{-2q}w(r/z)$ as expected, where $w(r/z)$ must solve the corresponding Eq. (10). In other words, these solutions exhibit the peculiar property of being simultaneously invariant under the "standard" symmetry (5) and the weak conditional one (19). A similar property holds for the solution (18), which turn out to be invariant under the standard symmetry (5) (with $q=-\gamma$), and under the conditional symmetry (17).

As a first explicit example of Eqs. (21) and (22), assuming, e.g., $\sigma=2$, if

$$F = -\alpha^2/\psi^3, \quad G = 0$$

i.e., with $q=-1/2$, $b=0$, we obtain the solution (which has been found in Ref. 20)

$$\psi(r,z) = \frac{\alpha^{1/2}}{2^{1/4}}(2r^2 + z^2)^{1/2}.$$

This corresponds to the solution

$$w(y) = \frac{\alpha^{1/2}}{2^{1/4}}\left(\frac{2y^2 + 1}{y^2}\right)^{1/2} \quad (y = r/z)$$

of the Eq. (10).

### A. The "doubling of solutions"

It can be emphasized that the coefficient $\sigma^2 - \sigma$ of $\psi_{ss}$ in the first equation in Eq. (20) takes the same value also replacing $\sigma=2$ with $\sigma=-1$, which gives $s=z^2-r^2$. No harm in evaluating square roots of quantities which may be negative, it is easy indeed to verify that in this case also

$$\psi(r,z) = \frac{\alpha^{1/2}}{2^{1/4}}|r^2 - z^2|^{1/2}$$

is another solution of the above equation.

This doubling of solutions is clearly true also in general: if $\sigma_1 + \sigma_2 = 1$, then the two solutions

$$\psi_1(r,z) = \psi(\sigma_1 r^2 + z^2) \quad \text{and} \quad \psi_2(r,z) = \psi(\sigma_2 r^2 + z^2)$$

hold simultaneously. This is an example of a "partial" discrete symmetry: i.e., a symmetry which holds only in a well defined subset of solutions. The invariant solutions under this discrete symmetry correspond to the choice $\sigma_1 = \sigma_2 = 1/2$, and then to solutions of the form $\psi(r,z) = \psi(r^2/2 + z^2)$.

Some other examples can be easily given: still with $\sigma=2$ or $\sigma=-1$, we get the trivial solution $\psi = z^2 - r^2$ if $F=0$, $G=2$; choosing for instance

$$F(\psi) = \pm \alpha^2, \quad G(\psi) = \pm 3\alpha|\psi|^{1/2}$$

we get the solutions

$$\psi(r,z) = \pm \frac{\alpha^2}{16}(2r^2 + z^2)^2 \quad \text{and} \quad \psi(r,z) = \pm \frac{\alpha^2}{16}(r^2 - z^2)^2.$$

If

$$F(\psi) = \pm \alpha^2 \psi^{1/3}, \quad G(\psi) = \pm \frac{15\alpha}{2\sqrt{3}}\psi^{2/3}$$

a solution is

$$\psi = \pm \frac{\alpha^2}{192\sqrt{3}}(2r^2 + z^2)^3.$$

### B. The D-shaped equilibrium solutions

We now consider in some detail the solutions with $q=-1/4$, i.e.,

$$\psi(r,z) = [A(\sigma r^2 + z^2)]^{1/4}$$

with $\sigma=2$ this solution holds without restrictions; the same is clearly true with any $\sigma>0$. Choosing instead $\sigma<0$, e.g., $\sigma=-1$, we get two different possibilities: the most interesting is the one with $A<0$, e.g., $A=-1$, i.e.,

$$\psi(r,z) = (r^2 - z^2)^{1/4}$$

which solves the GS equation with $F=-(3/2)\psi^{-7}$, $G=(1/4)\psi^{-3}$, according to Eq. (23), and clearly holds only in the region $|z| \leq r$. Taking now advantage from the "exceptional" symmetry $X''$ [Eq. (7)] which is present in this case, we can construct from this solution a continuous family of solutions using the rule (11): we obtain, for generic $\sigma<0$ and $A<0$

$$\psi(r,z) = [A\sigma r^2 - |A|(z^2 + 2\lambda z(r^2+z^2) + \lambda^2(r^2+z^2)^2)]^{1/4} \quad (24)$$

which holds in the interior of the two circles centered resp. in $r_0=\sqrt{|\sigma|}/(2\lambda)$, $z_0=-1/(2\lambda)$, and $r_0=-\sqrt{|\sigma|}/(2\lambda)$, $z_0=-1/(2\lambda)$, both of radius $\sqrt{1+|\sigma|}/(2\lambda)$, excluding their intersection (it is not restrictive to assume $\lambda>0$). A GS equation with $F=a\psi^{-7}$, $G=b\psi^{-3}$ is solved by the above flux function if

$$A = -4b, \quad \sigma = \tfrac{1}{2}(1-\sqrt{1-a/3b^2}) < 0$$

according to Eq. (23). Notice that the condition $a \leq 3b^2$ is equivalent to the condition $\sigma^2 - \sigma + 1/4 \geq 0$ ensuring the existence of a real parameter $\sigma$ and therefore of the variable $s$, which is the coordinate invariant under the weak symmetry $Y$ [Eq. (19)].

Thanks to the invariance of the GS equation under translations $z \to z+$const., we can shift the coordinate $z_0$ defined below Eq. (24) to 0. The resulting D-shaped configurations are shown in Fig. 3 for two different values of $\sigma<0$. The corresponding expressions for the pressure function $p(\psi)=p(r,z)$ and of the azimuthal magnetic field function $I(\psi)=I(r,z)$ are given by

$$p(r,z) = p_0 + \frac{a}{24\pi}\frac{1}{\psi^6}, \quad I^2(r,z) = I_0^2 + \frac{b}{\psi^2}.$$

In Fig. 4 the three-dimensional (3D) plots of the flux function $\psi(r,z)$, of the pressure $p(r,z)=(1/16\pi)[\psi_0^{-6}-\psi^{-6}]$, and of the azimuthal magnetic field function $I(\psi)=I(r,z)= \pm 1/(2\psi)$ corresponding to the solution of Eq. (24) with $\lambda=1$ and $I_0=0$ are shown in the domain $\psi > \psi_0=0.4$. At $\psi=\psi_0$ (plasma border) the pressure vanishes. This solution exhibits a rather flat plasma profile with strong pressure gradients and current gradients at the plasma edge. The corresponding safety parameter $q(\psi)$, defined as the derivative of the azimuthal field flux with respect to $\psi$, ranges in this case approximately from 1 to 5, increasing monotonically from the magnetic axis outward.

The other possibility, still with $\sigma=-1$ but $A>0$, e.g., $A=1$, is

$$\psi(r,z) = (-r^2+z^2)^{1/4}$$

which holds for $|z| \geq r$ and solves the GS equation with $a=-3/2$, $b=-1/4$. Using the same procedure as before, we obtain the family of solutions

$$\psi = [-r^2+z^2 + 2\lambda z(r^2+z^2) + \lambda^2(r^2+z^2)^2]^{1/4} \quad (25)$$

which holds in the complementary region defined for the previous case.

As a final remark, observing that the Lie commutator of the symmetries $X_1$ and $X''$ [resp. Eqs. (5) and (7)] is given by

$$[X_1, X''] = X''$$

we can conclude that all solutions invariant under $X''$ are mapped by the transformations generated by $X_1$ into solutions which are still invariant under $X''$. In particular, the simple solution $\psi \propto \sqrt{r}$ is invariant both under $X''$ and under $X_1$.

## V. CONCLUSION

An exhaustive review has been presented of the possible choices for the functions $F(\psi)$ and $G(\psi)$ that yield GS equations admitting some continuous Lie symmetries; some solutions related to these symmetries (in particular symmetry-invariant solutions) have been obtained and discussed.

Most remarkably, the introduction of conditional (and weak conditional) symmetries in the context of the solutions of the GS equation has allowed us to find a new family of solutions that correspond to D-shaped toroidal plasma equilibria with sharp gradients at the plasma edge. In general, finding the (weak) conditional symmetries of a differential equation in a systematic or algorithmic way is very difficult. Therefore the weak conditional symmetry found here and leading to the new family of solutions need not exhaust the possible useful choice of conditional and weak conditional symmetries in the context of the GS equation.

In conclusion, it should be emphasized the crucial role played by symmetry properties, in their different aspects, in the problem of finding solutions of differential equations.

### ACKNOWLEDGMENTS

We wish to thank Professor Philip J. Morrison for bringing Ref. 6 to our attention.

[1] P. J. Olver, *Application of Lie Groups to Differential Equations* (Springer, Berlin, 1986).
[2] *CRC Handbook of Lie Group Analysis of Differential Equations* (3 vols.), edited by N. H. Ibragimov (CRC, Boca Raton, 1994).
[3] G. W. Bluman and S. C. Anco, *Symmetry and Integration Methods for Differential Equations* (Springer, New York, 2002).
[4] G. Cicogna, F. Ceccherini, and F. Pegoraro, Symmetry, Integr. Geom.: Methods Appl. **2**, paper 017 (2006).
[5] G. Cicogna, Nonlinear Dyn. **51**, 309 (2007).
[6] M. Frewer, M. Oberlack, and S. Guenther, Fluid Dyn. Res. **39**, 647 (2007).
[7] R. H. White and R. D. Hazeltine, Phys. Plasmas **16**, 123101 (2009).
[8] L. S. Solovèv, in *Reviews of Plasma Physics*, edited by M. A. Leontovich (Consultants Bureau, New York, 1976), Vol. 6, p. 239.

[9]S. C. Cowley, P. K. Kaw, R. S. Kelly, and R. M. Kulsrud, Phys. Fluids B **3**, 2066 (1991).
[10]R. Y. Neches, S. C. Cowley, P. A. Gourdain, and J. N. Leboeuf, Phys. Plasmas **15**, 122504 (2008).
[11]C. V. Atanasiu, S. Gunter, K. Lackner, and I. G. Miron, Phys. Plasmas **11**, 3510 (2004).
[12]P. J. Mc Carthy, Phys. Plasmas **6**, 3554 (1999).
[13]S. B. Zheng, A. J. Wootton, and E. R. Solano, Phys. Plasmas **3**, 1176 (1996).
[14]D. Levi and P. Winternitz, J. Phys. A **22**, 2915 (1989).
[15]*Symmetry Analysis of Equations of Mathematical Physics*, edited by W. I. Fushchych (Institute of Mathematics Academy of Science of Ukraine, Kiev, 1992); *Modern Group Analysis: Advanced Analytical and Computational Methods in Mathematical Physics*, edited by N. H. Ibragimov, M. Torrisi, and A. Valenti (Kluwer, Dordrecht, 1993), p. 231.
[16]P. J. Olver, and Ph. Rosenau, Phys. Lett. A **114**, 107 (1986); SIAM J. Appl. Math. **47**, 263 (1987).
[17]G. Cicogna and M. Laino, Rev. Math. Phys. **18**, 1 (2006).
[18]F. Ceccherini, G. Cicogna, and F. Pegoraro, J. Phys. A **38**, 4597 (2005).
[19]M. Kunzinger and R. O. Popovych, *Proceedings of the 4th Workshop Group Analysis of Differential Equations and Integrability*, edited by N. Ivanova, C. Sophocleous, R. Popovych, P. Damianou, and A. Nikitin (University of Cyprus, Nicosia, 2008), p. 107.
[20]A. H. Khater and S. M. Moawad, *Proceedings of the IAU Symposium No. 233*, edited by V. Bothner and A. A. Hady (Cambridge University Press, Cambridge, U.K., 2006), p. 307.